\documentclass[final,1p,times]{elsarticle} 
\usepackage{graphicx}
\usepackage{amssymb} 
\usepackage{amsthm} 
\usepackage{amsmath} 
\usepackage{lineno}
\usepackage{mciteplus}
\usepackage{wrapfig}
\usepackage{hyperref}

\newcommand{\qs}{Q_\mathrm{s}}

\newcommand{\nc}{{N_\mathrm{c}}}

\newcommand{\ntt}{n_T} % scalar
\newcommand{\xt}{\mathbf{x}_T}
\newcommand{\yt}{\mathbf{y}_T}

\newcommand{\ut}{\mathbf{u}_T}
\newcommand{\vt}{\mathbf{v}_T}

\newcommand{\gev}{\textrm{ GeV}}

\newcommand{\fig}{fig.~}

\newcommand{\re}{ref.~}
\newcommand{\res}{refs.~}

\newcommand{\qso}{Q_\mathrm{s0}}

\journal{Nuclear Physics A} 

\begin{document}

\begin{frontmatter} 

% Your Title - please insert
\title{Forward particle correlations in the color glass condensate}

%% Single author (and collaboration) - please insert
\author[jyfl,hip]{T. Lappi}
\author[jyfl]{H. M\"antysaari}
\address[jyfl]{Department of Physics, %
 P.O. Box 35, 40014 University of Jyv\"askyl\"a, Finland}
\address[hip]{Helsinki Institute of Physics, P.O. Box 64, 00014 University of Helsinki,Finland}

\begin{abstract} 
 Multiparticle correlations, such as forward dihadron correlations 
in pA collisions, are an important probe of
  the strong color fields that dominate the initial stages of a heavy ion 
 collision. 
We describe recent progress in 
understanding two-particle correlations in the dilute-dense system, 
e.g. at forward rapidity  in deuteron-gold collisions. 
This requires evaluating higher point Wilson
 line correlators from the JIMWLK equation, which we find well
described by a   Gaussian  approximation.
We then calculate the  dihadron correlation, including both the
``elastic'' and ``inelastic'' contributions, and show
that our result includes the double parton scattering contribution.
\end{abstract} 

\end{frontmatter} % do not change

\section{Introduction}\label{sec:intro}
The physics of high energy hadronic or nuclear collisions is dominated
by the gluonic degrees of freedom of the colliding particles. 
These
small~$x$ gluons form a dense nonlinear system that is, at high enough
$\sqrt{s}$, best described as a classical color field and quantum 
fluctuations around it. 
The color glass condensate (CGC,
for  reviews see \cite{Iancu:2003xm,*Weigert:2005us,*Gelis:2010nm,*Lappi:2010ek})
is an effective theory developed around this idea. It gives an 
universal description of the small~$x$ degrees of freedom 
that can equally well be applied to small~$x$
DIS as to  dilute-dense (pA or forward AA) and dense-dense (AA or very 
high energy pp) hadronic collisions. 
The nonlinear interactions of the small $x$ gluons dynamically generate  
a new transverse momentum scale, the saturation scale $\qs$,
that grows with energy.
The scale $\qs$ dominates both the
gluon spectrum and multiparton correlations.

The most convenient parametrization of the dominant gauge field
is in terms of Wilson lines that describe the eikonal propagation 
of a projectile through it. 
The Wilson lines are drawn from a probability distribution, whose
dependence on rapidity is described by the JIMWLK equation.
This equation reduces, in a large
$\nc$ and mean field approximation, to the 
BK~\cite{Balitsky:1995ub,*Kovchegov:1999yj} equation and further, in the
dilute linear regime, to the BFKL one.

\section{Correlations in a dilute-dense collision}\label{sec:multipt}

\begin{figure}
\includegraphics[width=0.5\textwidth]{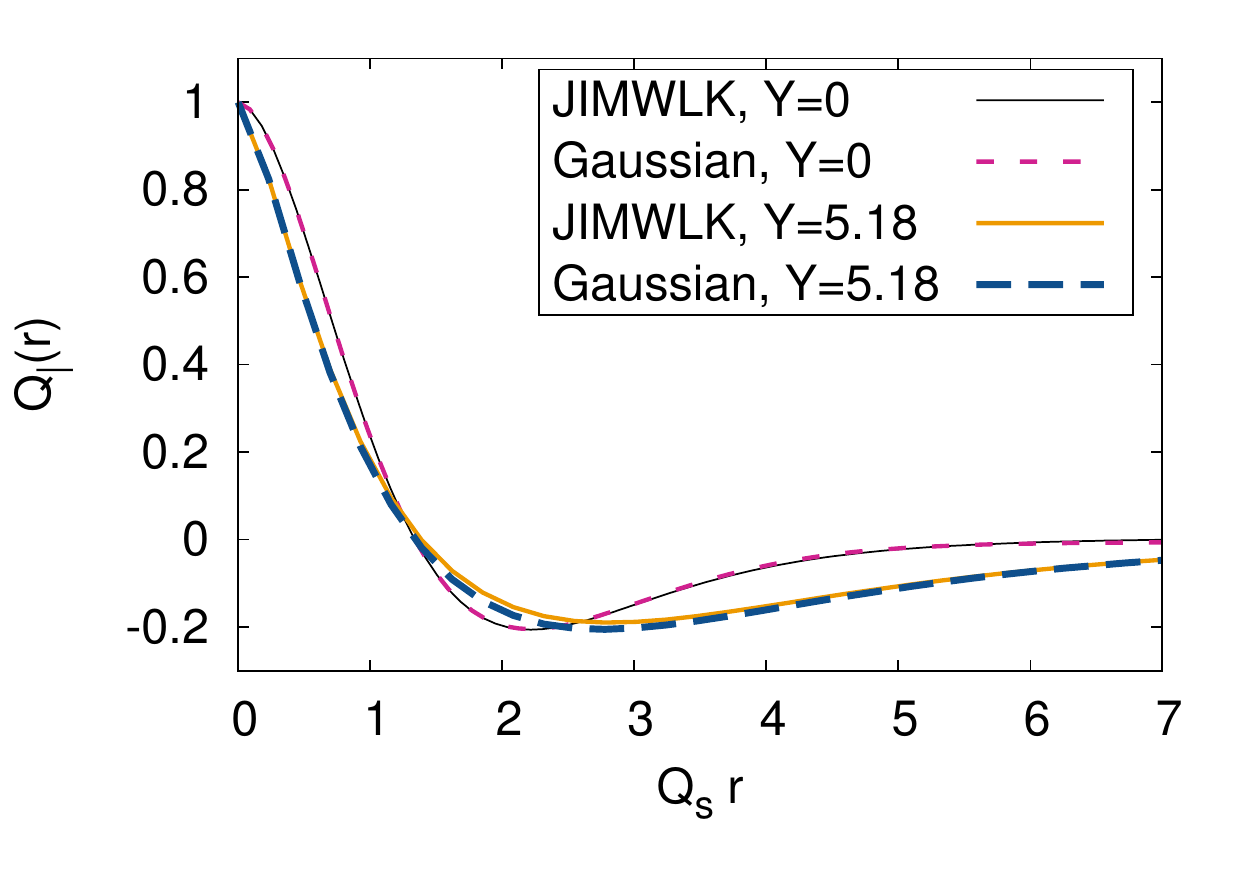}
\includegraphics[width=0.5\textwidth]{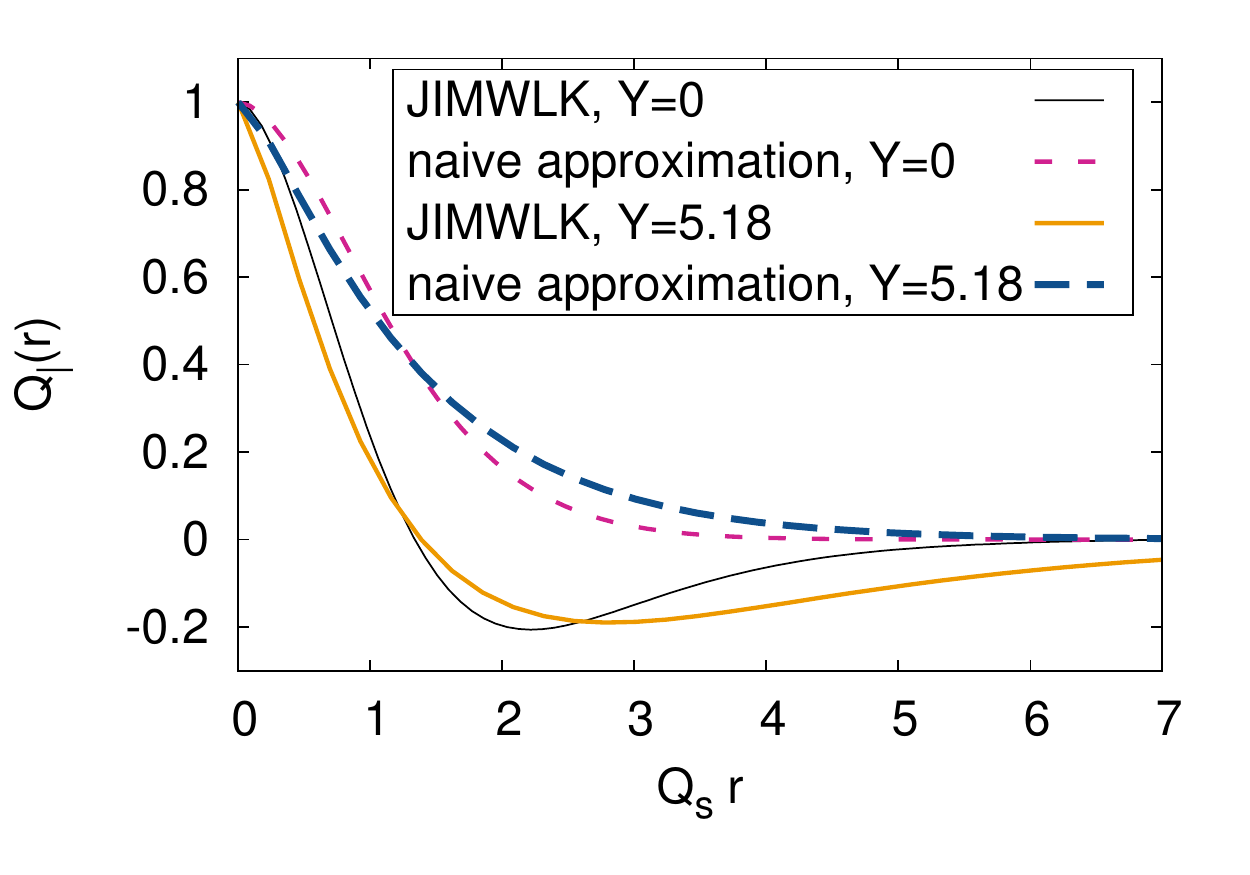}
\caption{\label{fig:4pt}
Left: the JIMWLK result for the quadrupole correlator 
compared to the
Gaussian approximation. Right: Comparison
 to the ``naive large $\nc$'' approximation.
 Shown are the initial condition (MV model) at 
$y=0$ and the result after $5.18$ units of evolution 
in rapidity in the ``line'' coordinate configuration
where $\xt=\ut$ and $\yt=\vt$.
Figures from \re\cite{Dumitru:2011vk}.
}
\end{figure}

One of the more striking signals of  saturation physics
at RHIC is 
seen in the relative azimuthal angle ($\Delta \varphi$) dependence 
of the dihadron correlation function, where the 
$\Delta \varphi\approx \pi$ back-to-back peak is seen to 
be suppressed in dAu-collisions compared to pp collisions at
the same kinematics~\cite{Adare:2011sc,Braidot:2011zj}.
The CGC description of this correlation starts from a large~$x$
parton radiating a gluon, with subsequent
 eikonal  propagation of the pair through the target. 
To calculate the matrix element for this process 
one needs target expectation values of  products of Wilson line
operators, such as the dipole and the quadrupole
\begin{equation}\label{eq:dipquad}
\hat{D}(\xt,\yt) = 
\frac{1}{\nc} \mathrm{Tr} U(\xt) U^\dagger(\yt)
\quad 
\hat{Q}(\xt,\yt,\ut,\vt) = 
\frac{1}{\nc} \mathrm{Tr} U(\xt) U^\dagger(\yt)
U(\ut) U^\dagger(\vt).
 \end{equation}
For practical phenomenological work it would be extremely convenient to be 
able to express these higher point correlators in terms of the dipole,
which is straightforward to obtain from the BK equation.
In the phenomenological literature so 
far~\cite{Albacete:2010pg} this has been done using
a ``naive large $\nc$'' (or ``elastic'') approximation where the quadrupole is assumed 
to be simply a product of two dipoles.
A more elaborate scheme would be a ``Gaussian'' approximation
(``Gaussian truncation'' in~\cite{Kuokkanen:2011je}), where
one assumes the relation between the higher point functions and the
dipole to be the same as in the (Gaussian) MV model. The 
expectation value of the quadrupole operator
in the MV model has been derived e.g. in \re\cite{Dominguez:2011wm}.

In \re\cite{Dumitru:2011vk} the validity of these approximations 
was studied by comparing them numerically to the 
solution of the JIMWLK equation.
As studying the full 8-dimensional phase space for the quadrupole operator
would be cumbersome, the numerical study was done in   two
special coordinate configurations.
The most important results of \re\cite{Dumitru:2011vk} 
 is demonstrated in Fig.~\ref{fig:4pt},
with a comparison of the initial and evolved (for 5.18 units in $y$)
JIMWLK results to the approximations.
The MV-model initial condition $y=0$ satisfies the Gaussian approximation 
by construction, but the calculation shows that the Gaussian approximation
is still surprisingly well conserved by the evolution\footnote{A possible explanation
for the success of the Gaussian approximation has  
been proposed in~\cite{Iancu:2011ns,*Iancu:2011nj}.}. 
The naive large  $\nc$ approximation,
on the other hand,  fails already at the initial condition.

\section{Dihadron correlation}

This result does not yet fully address the effect on the
measurable dihadron cross section.
For that one must convolute a linear combination of Wilson line operators
$\langle\hat{Q}\hat{D} \rangle,\langle\hat{D}\hat{D} \rangle$ and $\langle\hat{D} \rangle$
with the $q\to qg$ splitting wavefunction. For the explicit expressions we refer the reader to
\res\cite{Albacete:2010pg,Lappi:2012nh}. We have reported the outcome of this  nontrivial
numerical task in more detail in \re\cite{Lappi:2012nh}.  

\begin{wrapfigure}{R}{0.4\textwidth}
\centerline{\includegraphics[width=0.4\textwidth]{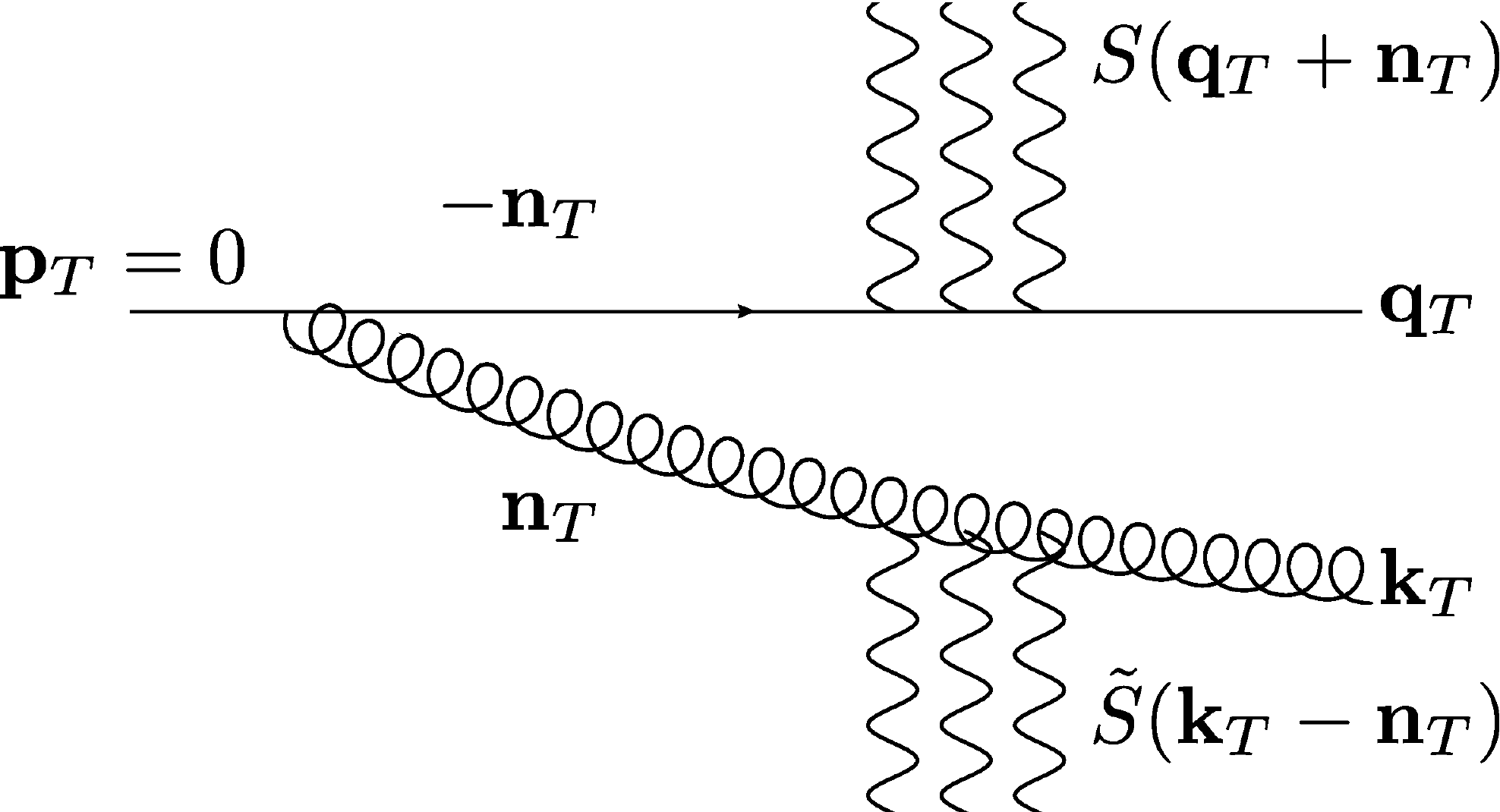}}
\caption{\label{fig:dpskin}
Kinematics of the DPS limit. The double parton scattering contribution is obtained from
the CGC dihadron cross section formula in the limit $\ntt\to0$.
}
\end{wrapfigure}
 
When using the full Gaussian correlator, instead of only the ``elastic'' term, one encounters
an additional complication in the calculation, not encountered with the approximations
used in \res\cite{Albacete:2010pg,Stasto:2011ru}. In the limit of large distance, or small momentum,
between the quark and the gluon, the operator
$\langle\hat{Q}\hat{D} \rangle$ factorizes into a product of an adjoint and a fundamental 
representation dipole operator,  describing the independent scattering of the
quark and the gluon, respectively, off the target (see \fig\ref{fig:dpskin}). 
When this product is multiplied by the splitting wavefunction, the resulting integral 
is logarithmically infrared divergent. This divergence
corresponds to a correlated quark-gluon pair being present in the incoming probe wavefunction, 
namely the double parton scattering (DPS) contribution 
discussed in this context in \re\cite{Strikman:2010bg}. For a consistent treatment it
must be subtracted from the correlated cross section and calculated separately using an additional
nonperturbative input describing the probe, a double parton distribution function.

\begin{figure}[b!]
\begin{minipage}[t]{0.48\linewidth}
\includegraphics[width=\textwidth]{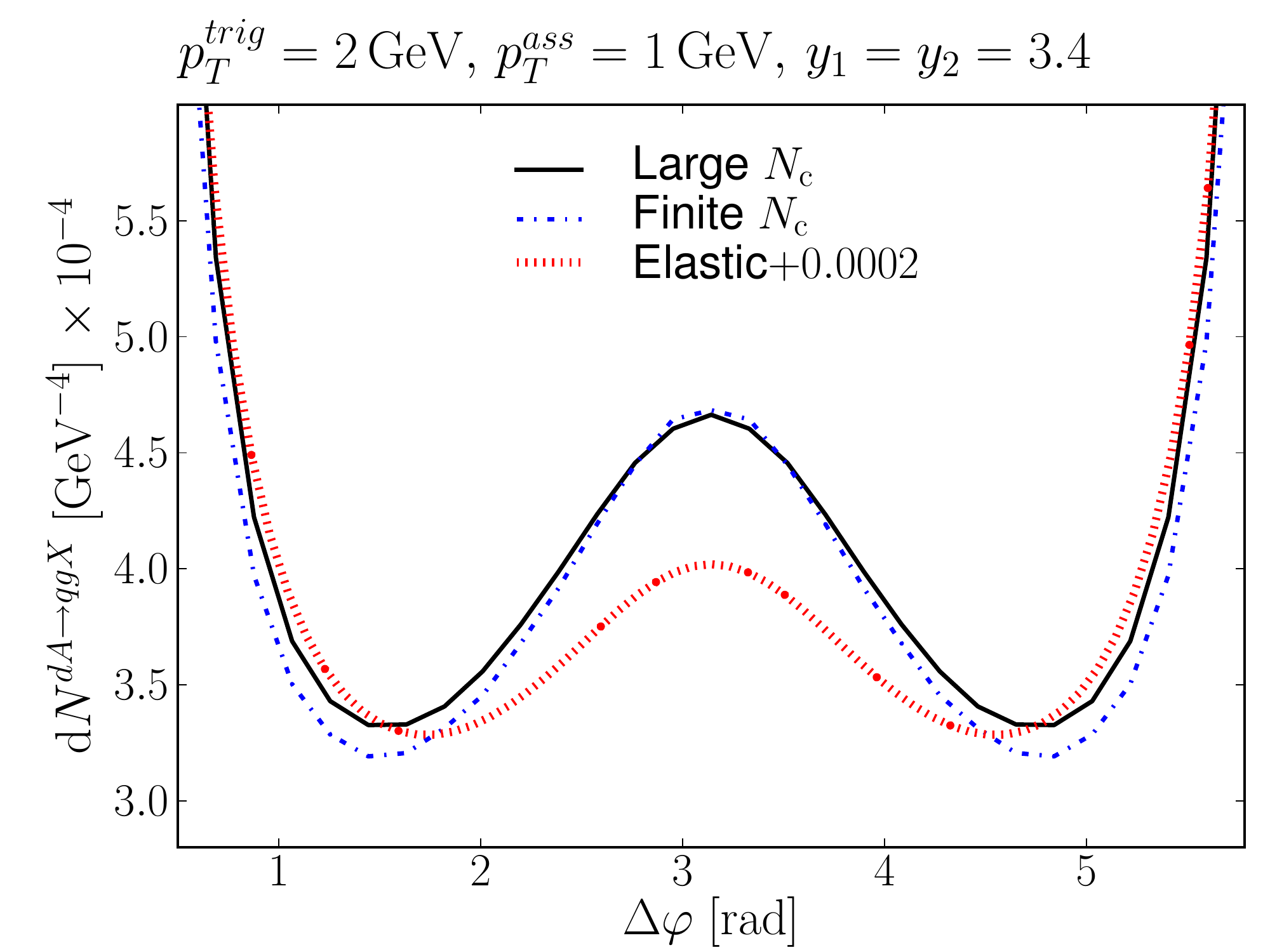}
\caption{\label{fig:dihad}
A parton level comparison of the elastic approximation to our Gaussian approximation
and its large $\nc$ limit.
}
\end{minipage}
\hfill
\begin{minipage}[t]{0.48\linewidth}
\includegraphics[width=\textwidth]{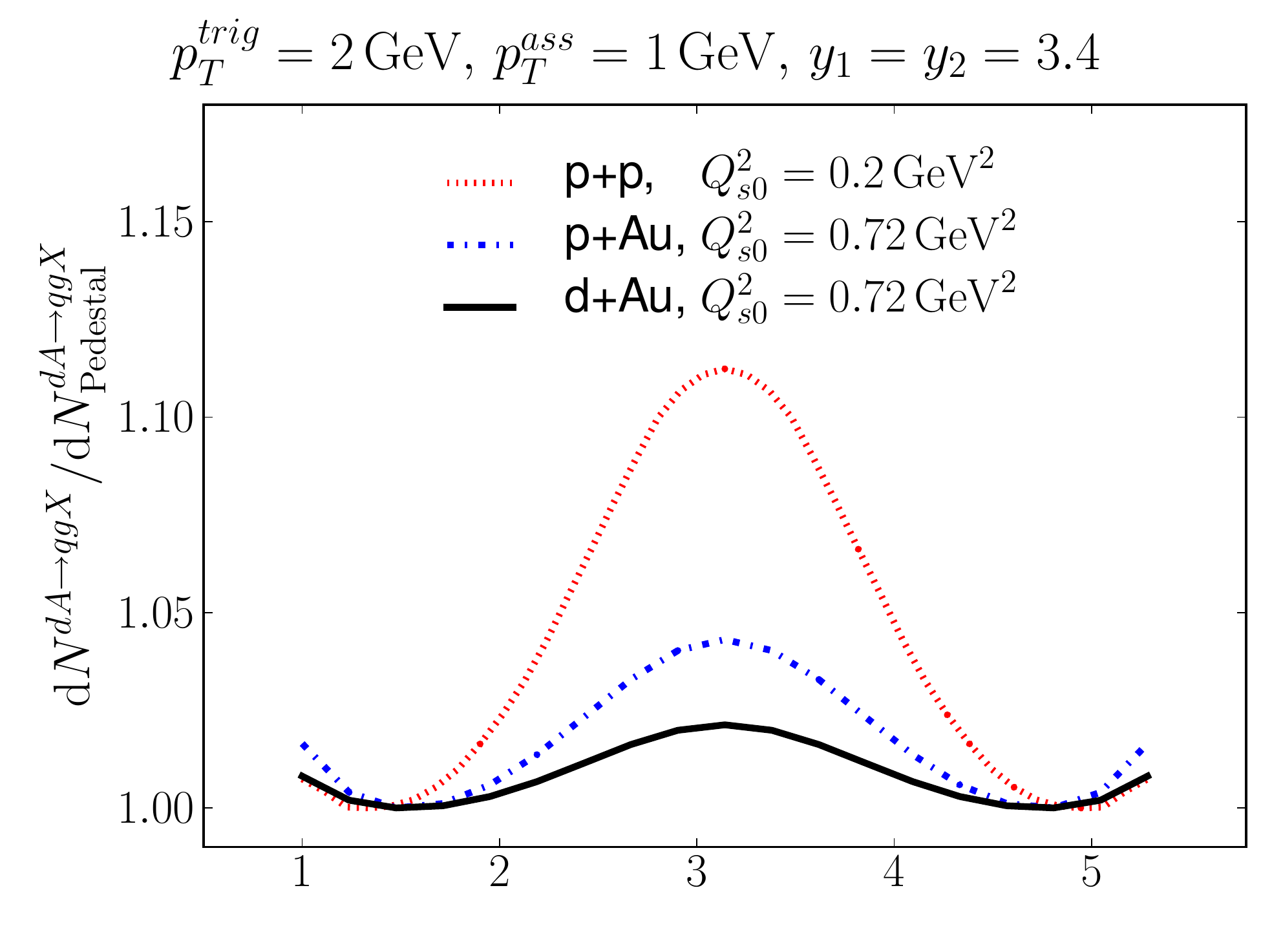}
\caption{\label{fig:dps}
Illustration of the effect of 
the DPS contribution on the parton level:
 the dihadron yield divided by the $\Delta \varphi$-indpendent part.
}
\end{minipage}
\end{figure}

In \fig\ref{fig:dihad} we show the parton level dihadron production cross section obtained 
using different approximations for the quadrupole.
We notice that the width and especially the height of the away side peak are modified
when replacing the naive approximation, used e.g. in \cite{Albacete:2010pg}, by the
Gaussian approximation \cite{Dominguez:2011wm}. In \fig\ref{fig:dps} we show the yield  
divided by the $\Delta \varphi$-independent pedestal for pp, pAu and dAu collisions. The
back-to-back correlation is suppressed  for a nuclear target, due to the larger intrinsic
transverse momentum $\sim \qs$. When the probe is switched from a proton to a deuteron,
on the other hand, the correlated peak stays the same but the $\Delta\varphi$-independent
background increases, resulting in  a decreased ratio of the peak to the pedestal, shown 
in the plot. This is the effect discussed in much detail in \re\cite{Strikman:2010bg}.

\begin{figure}
\includegraphics[width=0.5\textwidth]{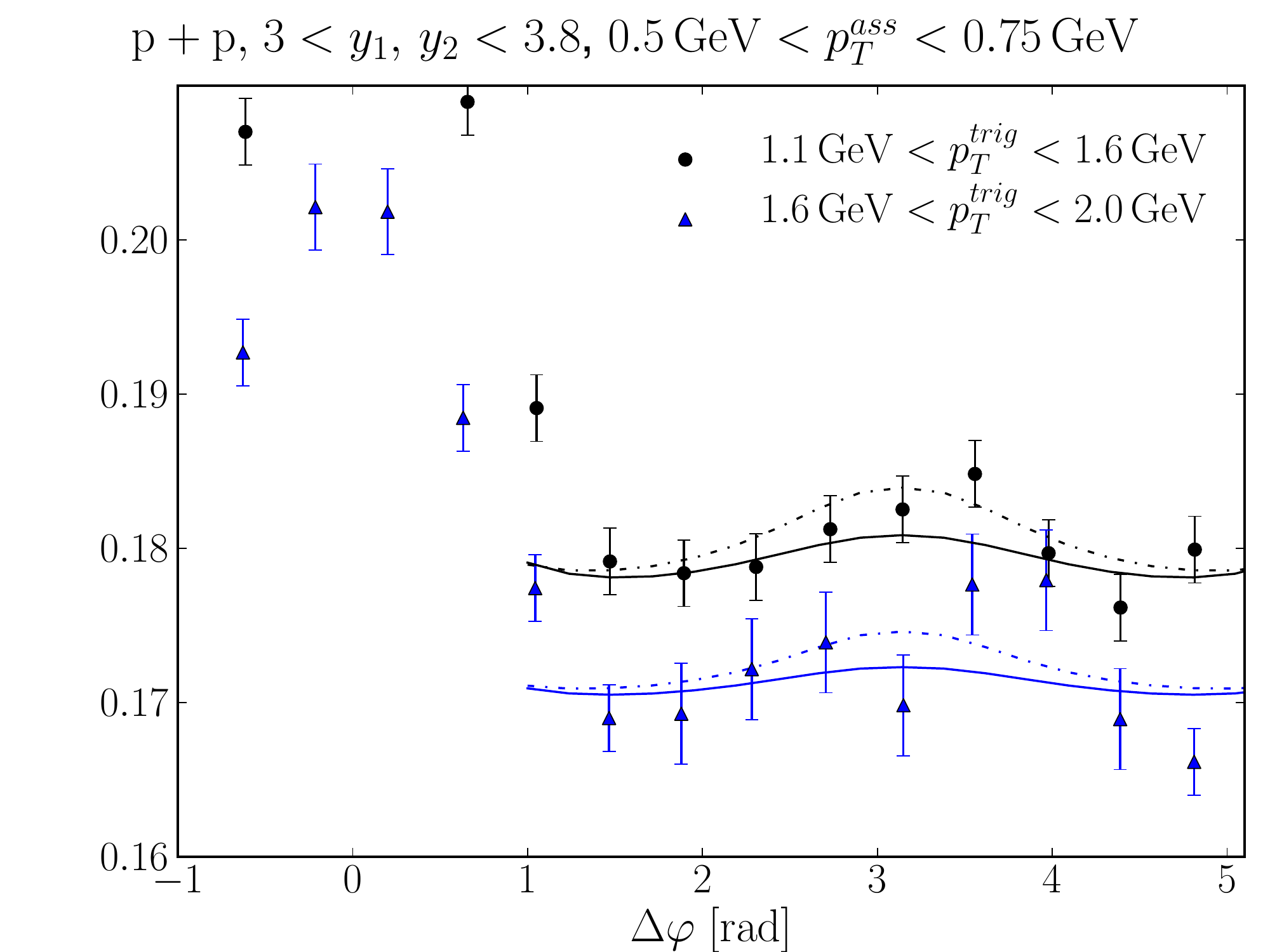}
\includegraphics[width=0.5\textwidth]{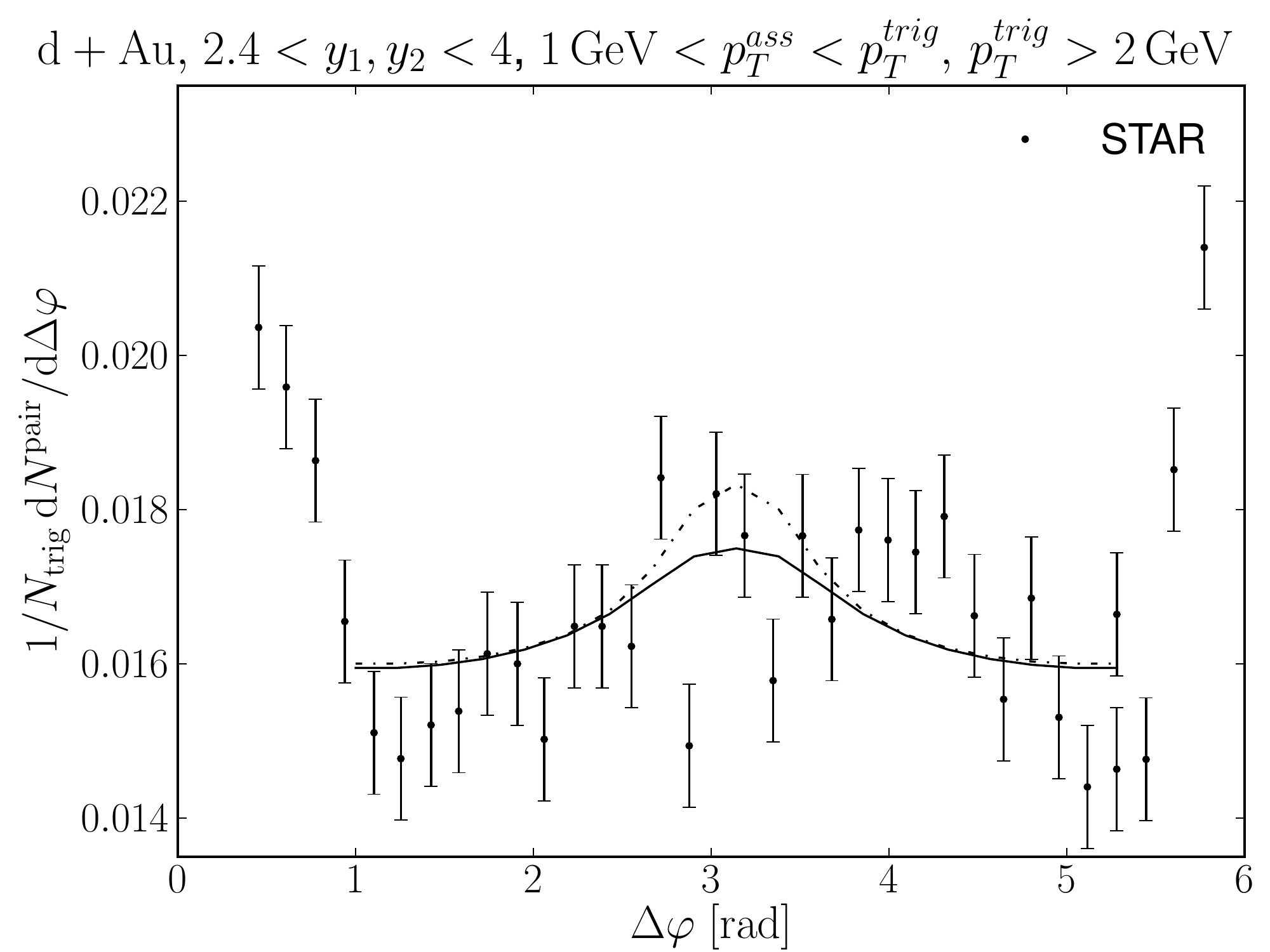}
\caption{\label{fig:expdata}
Comparison of our calculation to the PHENIX (left, \cite{Adare:2011sc})  and
STAR (right,~\cite{Braidot:2011zj})
dihadron correlation data. The initial saturation scales are $\qso^2=1.51 \gev^2$ (solid line) and
$\qso^2=0.72 \gev^2$ (dashed line).
}
\end{figure}

There is some remaining  uncertainty in the normalization of the
single inclusive spectrum (as evidenced by 
substantial  K-factors needed in the literature to describe the single inclusive data),
which propagates to a factor $\sim 2$ uncertainty in our estimates for the
$\Delta \varphi$-independent pedestal. 
When comparing with PHENIX~\cite{Adare:2011sc} data, we obtain for the trigger transverse 
momentum range $1.1\dots 1.6 \gev$
a pedestal $0.11 \gev^{-1}$, whereas the experimental value is $0.176 \gev^{-1}$.
Similarly for the trigger transverse momentum $1.6 \dots 2\gev$ we obtain $0.08 \gev^{-1}$, and
 the experimental value reads $0.163 \gev^{-1}$. For the 
STAR~\cite{Braidot:2011zj} data our estimate for the pedestal is $0.02$ when the experimental value
is $0.0145$. In order to compare to the experimental correlation
peaks from the PHENIX and 
 experiments we have adjusted this pedestal to the data. 
The resulting comparison  is shown in \fig \ref{fig:expdata}.

\paragraph{Acknowledgements}
H.M. is supported by the Graduate School of Particle and Nuclear Physics.
This work  has been supported by the Academy of Finland, project
133005 and by computing resources from
CSC -- IT Center for Science in Espoo, Finland.

\bibliographystyle{h-physrev4mod2M}
\bibliography{spires}

\end{document}